\documentclass[fleqn,usenatbib]{mnras}  
\pdfoutput=1
\usepackage{newtxtext,newtxmath}
\usepackage[export]{adjustbox}
\usepackage{threeparttable}
\usepackage{caption}
\usepackage[T1]{fontenc}
\usepackage{ae,aecompl}
\usepackage{natbib}

\usepackage{graphicx}	
\usepackage{amsmath}	
\usepackage{amssymb}	
\usepackage{diagbox}    

\title[Quasar variability]{Solving the puzzle of discrepant quasar variability on monthly time-scales implied by SDSS and CRTS data sets}

\author[K. Suberlak et al.]{
Krzysztof Suberlak,$^{1}$\thanks{E-mail: suberlak@uw.edu}
\v{Z}eljko Ivezi\'c,$^{1}$
Chelsea L. MacLeod,$^{2}$
Matthew Graham$^{3,4}$
\newauthor
$\, \,  $and Branimir Sesar$^{5}$
\\
$^{1}$Department of Astronomy, University of Washington, Seattle, WA 98195, USA\\
$^{2}$Harvard-Smithsonian Center for Astrophysics, Cambridge, MA 02138, USA\\
$^{3}$Center for Data-Driven Discovery, California Institute of Technology, Pasadena, CA 91125, USA\\
$^{4}$National Optical Astronomy Observatory, Tucson, AZ 85719, USA\\
$^{5}$Max Planck Institute for Astronomy, K\"{o}nigstuhl 17, D-69117 Heidelberg, Germany. 
}

\date{Accepted 2017 September 4. Received 2017 August 17; in original form 2017 July 5}

\pubyear{2017}

\begin{document}
\label{firstpage}
\pagerange{\pageref{firstpage}--\pageref{lastpage}}
\maketitle

\begin{abstract}
We present an improved photometric error analysis for the 7,100 CRTS (Catalina Real-Time Transient Survey)  optical  light curves for quasars from the SDSS (Sloan Digital Sky Survey) Stripe 82 catalogue. The SDSS imaging survey  has provided a time-resolved photometric  data set which greatly improved our understanding of the quasar optical continuum variability: Data for monthly and longer time-scales  are consistent with a damped random walk (DRW). Recently, newer data  obtained by CRTS provided  puzzling evidence for enhanced variability, compared to SDSS  results, on monthly time-scales. Quantitatively, SDSS results predict  about 0.06 mag root-mean-square (rms) variability  for monthly time-scales, while CRTS data show about a factor of 2 larger rms, for spectroscopically confirmed SDSS quasars. Our analysis has successfully resolved this discrepancy as due to slightly underestimated photometric uncertainties from the CRTS image processing pipelines. As a result, the correction for observational noise is too small and the implied quasar variability is too large. The CRTS photometric error correction factors, derived from detailed analysis of non-variable SDSS standard stars that were re-observed by CRTS, are about 20-30\%, and result in reconciling  quasar variability behaviour implied by the CRTS data with earlier SDSS results. An additional analysis based on independent light curve data for the same objects obtained by the Palomar Transient Factory provides further support for this conclusion. In summary, the quasar variability constraints on weekly and monthly time-scales from SDSS, CRTS and PTF surveys are mutually compatible, as well as consistent with DRW model.
\end{abstract}

\begin{keywords}
methods: data analysis -- techniques: photometric -- surveys -- quasars: general

\end{keywords}



\section{Introduction}
Variability can be used to both select and characterize quasars in sky surveys (for a recent overview see \citealt{lawrence2016a}). Although various time-scales of variability can be linked to physical parameters, such as accretion disc viscosity, or corona geometry (\citealt{kelly2011}; \citealt{graham2014}), the physical mechanism remains elusive. Most viable explanations for observed
variability include accretion disc instabilities \citep{kawaguchi1998}, surface thermal fluctuations from magnetic field turbulence \citep{kelly2009}, and
coronal X-ray heating \citep[][see \citealt{kozlowski2016} for a review]{kelly2011}.

The diversity of  physical scenarios available to explain the origin of quasar variability results in a variety of ways to characterize it. The two most widely used approaches to describing the variability of quasars include a structure function (SF) analysis and light curve
fitting based on damped random walk (DRW, also known as the Ornstein--Uhlenbeck process) model (\citealt{kelly2007}; \citealt{macleod2011}). An SF analysis essentially measures the 
width of the magnitude difference distribution as a function of the time separation, $\Delta t$. 
The DRW model approach is better suited for well-sampled light curves with a typical cadence of days \citep{zu2013, kozlowski2016}, whereas an ensemble SF analysis is better for sparsely sampled light curves \citep{hawkins2002, berk2004,  devries2005}; for a review and discussion see \cite{kozlowski2016}. Although the sampling for CRTS (the Catalina Real-time Transient Survey) light curves 
analysed here (see Section~\ref{sec:crtsdata}) might be adequate for light curve fitting, we nevertheless opt for 
the SF approach because it allows for more straightforward analysis when data quality is suspect. 

The observed SF is often characterized by a simple power law \citep{schmidt2010}. If the probed time-scales are long enough 
($\sim$ years), the power law flattens above a characteristic tim-scale, $\tau$ \citep{2004Ivezic, kelly2007, macleod2010}.
This time-scale may correspond to a transition from the stochastic thermal process that drives the variability to the physical 
response of the disc that successfully  dampens the amplitude on longer time-scales \citep{peterson2001, kelly2007, kelly2009, 
kelly2011, lawrence2016a}. In the context of a DRW model, the expected SF is described by 
\begin{equation}
\label{eq:DRWSF}
        {\rm SF}(\Delta t) = {\rm SF}_\infty \, \left[1 - \exp(-\Delta t / \tau) \right]^{1/2},
\end{equation}
where ${\rm SF}_\infty$ is the asymptotic value of the SF (for $\Delta t \ll \tau$, SF($\Delta t$) $\propto \Delta t^{1/2}$). 

Most studies found that $\tau > 100$ d (\citealt{macleod2010};  \citealt{kozlowski2016}). It is a relatively short time-scale compared to the dominant time-scale of variation for quasars, that exceeds 10 years \citep{hawkins2007}. Recently, \cite{graham2014} 
found a characteristic time-scale in quasar's rest frame of about 54 d, using the Slepian wavelet variance (SWV) analysis
of CRTS light curves (the SWV time-scale denotes the point at which the ensemble SWV for quasars deviates from
the ensemble SWV for a DRW  realization of the same data set, and is thus different from $\tau$ obtained in DRW analysis). 
This short time-scale implies much stronger variability on monthly time-scales than observed in SDSS 
data: SDSS results from \cite{macleod2010} predict about 0.06 mag root-mean-square (rms) variability for time-scales below 
50 d, while this CRTS-based analysis implies about a factor of 2 larger rms. These discrepancies have serious
implications for physical interpretations of quasar variability: Observed time-scales are directly related to physical processes
and increased variability levels call in question DRW as a viable model for describing quasar light curves 
(\citealt{macleod2010};  \citealt{kozlowski2016}). 

It is not obvious whether these discrepancies are due to various problems with the CRTS and/or SDSS data sets
(inadequate sampling, incorrect estimates of photometric errors, etc.), or perhaps are due to different analysis 
methods (SWV versus SF analysis). Here, we reanalyse these CRTS data using the same SF method as used by
\cite{macleod2010} to analyse SDSS data, and investigate the origin of these discrepant time-scales and 
variability levels. We argue that the most likely explanation of these discrepancies are slightly under-estimated
photometric errors for CRTS light-curve data.

\section{Data Sets}

We study stars and quasars selected from the sky region known as SDSS Stripe 82 (S82; an $\sim$300 deg$^2$ large
region along the celestial equator: $22^{h} 24^{m} < \mathrm{RA} < 04^{h} 08^{m}$ and $\mathrm{| Dec |} < 1.27^\circ$). 
We utilize both SDSS and CRTS photometric data. 

\subsection{Sloan Digital Sky Survey (SDSS)}

We use two SDSS catalogues, with five-band near-simultaneous photometry for 9258  quasars,  and 1$\,$006$\,$849 standard stars
(non-variable stars, as implied by the repeated SDSS photometry, see \citealt{ivezic2007}). 
The quasar catalogue\footnote{\url{http://www.astro.washington.edu/users/ivezic/cmacleod/qso_dr7/Southern.html}} includes spectroscopically confirmed quasars from the SDSS Data Release 7 \citep{abazajian2009}, based on the SDSS Quasar Catalogue V \citep{schneider2010}, and was compiled by \cite{macleod2012}. The SDSS standard stars
catalogue\footnote{\url{http://www.astro.washington.edu/users/ivezic/sdss/catalogs/stripe82.html}} 
was constructed as described in \cite{ivezic2007}.

\subsection{Catalina Real-time Transient Survey (CRTS) \label{sec:crtsdata}}
The main goal of CRTS was to find near-Earth objects. Its short intra-night cadence (four exposures per
night) was designed to allow a rapid follow-up \citep{graham2015b}, and white light (without filter) 
light curves maximize the sensitivity for faint objects. Three survey telescopes (the 0.7 m Catalina Sky Survey 
Schmidt in Arizona,  the 1.5 m Mount  Lemmon Survey telescope in Arizona, and the 0.5 m Siding Spring Survey 
Schmidt in Australia) were equipped with identical, 4kx4k CCDs (see \citealt{djorgovski2011a} for technical details).
Although, in principle, white light magnitudes can be calibrated to Johnson's V-band zero-point \citep{drake2013},
this step was unnecessary in our analysis. 

\begin{figure}
\vskip -0.2in
\includegraphics[width=1.04\columnwidth]{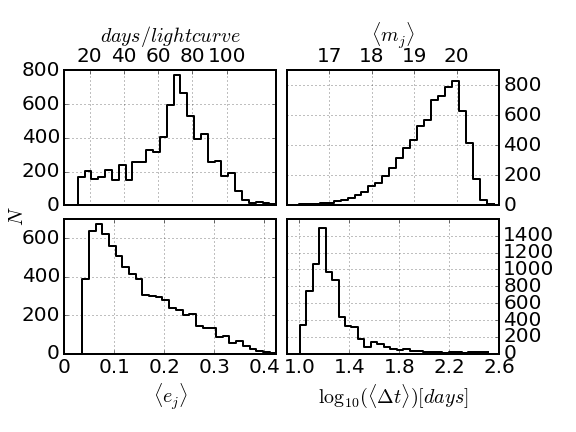}
\vskip -0.1in
\caption{The distribution of properties of 7601 CRTS quasar light curves for objects that were observed 
on at least 10 distinct nights (epochs). The  distribution of the number of distinct nights is shown in the 
upper left-hand panel. Within that sample, $96\%$ of light curves are longer than 7 years.  The upper right-hand panel 
shows the mean day-averaged CRTS magnitude, $\langle  m_{j} \rangle$ (see equation~\ref{eq:1}). 
The bottom left-hand panel shows the  mean day-averaged error, $\langle \sigma_{j} \rangle$ (see equation~\ref{eq:2}). 
We use only quasars with light curve averaged error smaller than 0.3, leaving 7,108 quasars in the sample.
The bottom right-hand panel shows the mean time difference $\langle \Delta t \rangle$ between day-averaged epochs. 
All means here are calculated per lightcurve. }
\label{fig:1}
\end{figure}

In this study, we used a sample of 7932 spectroscopically confirmed S82 quasars from the CRTS Data Release 2, based 
on the list by \cite{macleod2012}.  The majority (96\%) of  CRTS quasar light curves span the time of 7--9 yr, 
with typical sampling of 1--4 observations per night, 70 observing nights, on average, and the 
median interval between two successive observing nights is 17.52  d (see Fig.~\ref{fig:1}). We also use CRTS light curves for 52$\,$133 randomly chosen 10\% subsample of the S82 
standard stars from \cite{ivezic2007}. 

\subsection{Preprocessing}
\label{sec:preprocessing}
It is common to bin the data to reduce noise, by averaging over time-scales shorter than what is required by the science goals. In this study, the hourly time-scale of intra-night variability of CRTS light curves, with $\sim4$ epochs each night, is much shorter than the time-scales of interest (of the order of tens of days).  We day-averaged all CRTS light curves following a procedure similar to \cite{charisi2016}. We adopt a convention that an index  $i$ runs over intra-night observations, and an index $j$ separates distinct observing nights. Thus the day-averaged time-stamp is : 
\begin{equation}
t_{j} = \langle t_{ij} \rangle =  N^{-1} \, \sum_{i=1}^{N}{ t_{ij} }
\end{equation}
where $N$ is the number of observations per night. We similarly replace each set of $N$ brightness measurements from the  $j$-th  night by their mean weighted by the inverse square of error:
\begin{equation}
\label{eq:1}
 m_{j} = \langle m_{ij} \rangle = \frac{\sum_{i=1}^{N} {w_{i,j} m_{i,j}} } {\sum_{i=1}^{N} {w_{i,j}} }
\end{equation}
with weights $w_{i,j} = err_{i,j}^{-2}$, where $err_{i,j}$ are photometric uncertainty (colloquially, `error') estimates
for individual photometric data points computed by the CRTS photometric pipeline. Averaging in 
flux space, instead of magnitude space, would not qualitatively change the results (because photometric 
uncertainties are sufficiently small). 

Finally, we estimate the error on the weighted mean $m_{j}$ by the inverse square of the sum of weights:  
\begin{equation}
\label{eq:2}
err_{j} = \left(\sum_{i=1}^{N} {w_{i,j}}\right)^{-1/2}, 
\end{equation} 
and to avoid implausibly small error estimates, we add in quadrature 0.01 mag to $err_{j}$ if $err_{j} < 0.02$ mag
(note that for homoscedastic errors, $err_{i,j}=\overline{err}$, $err_j = \overline{err}/\sqrt{N}$).

\subsection{Final sample selection}

We have selected both quasars and stars using a combination of information from SDSS and CRTS. 
To find magnitude difference between different observing nights, we first require that the raw light curves
must have more than 10 photometric points (raw epochs). This step reduces the sample size from the initial 52,131 stars and 7,932 
quasars to 49,385 stars and 7,707 quasars. After day-averaging, we also remove light curves with less than 10 
observing nights (day-averaged epochs), leaving 48,250 stars and 7,601. In addition, we require that the light curve-average 
of nightly errors $\langle err_{j} \rangle < 0.3$ mag (see Fig.~\ref{fig:1}); this step removes
fewer than 10\% of light curves. Our final samples include 42,864 stars and 7,108 quasars. 

A crucial part of our analysis below is a test of photometric uncertainties computed by the CRTS photometric
pipeline using repeated CRTS observations of non-variable stars. In order to test for possible systematic
effects with respect to magnitude (most notably the increase of photometric noise towards
the faint end) and colour, we first select  subsamples from three magnitude bins, using the SDSS $r$ magnitudes: 
{\it bright:} 17-18,  {\it medium:} 18-18.5, and {\it faint:} 18.5-19. We note that the faint completeness limit 
of the SDSS spectroscopic quasar sample is $r\sim19$, and that the CRTS white light magnitudes are strongly
correlated with the SDSS $r$ magnitudes. Furthermore, we split the stellar sample using SDSS colour measurements
into the `blue' ($-$1$<g-i<$1) and `red' (1$<g-i<$3) subsamples. Table~\ref{tab:object_count} shows the 
number of objects in each type-magnitude bin.

\begin{table}
\centering
\begin{threeparttable}
\caption{Count of stars and quasars, selected by their SDSS $r$ magnitudes and $g$--$i$ colours.}
\label{tab:object_count}
\begin{tabular}{ lccc} 
\hline
$r$ magnitude & Red stars & Blue stars & Quasars \\ 
\hline
17-18   & 2993 & 2795   & 185    \\ 
18-18.5 & 2087 &  1400  & 333   \\ 
18.5-19 & 2327 &  1496  & 747   \\
& & & \\
Total   & 7407 &  5691 & 1265 \\
\hline
\end{tabular}
\end{threeparttable}
\end{table}

\section{Analysis}
\label{sec:analysis}

The structure function (SF) is a well-studied approach to characterizing light curves \citep{2004Ivezic, berk2004, devries2005, macleod2010, graham2013, kozlowski2016}. SF is closely related to the auto-correlation function (ACF), which in turn is the Fourier Transform of the  frequency power spectrum (PS) (for a detailed discussion,  see \citealt{ivezic2014,kozlowski2016}). We choose to analyse light curves with SF over PS because the main motivation for our paper is to resolve the discrepancy between quasar time-scales  found with SDSS data using the SF method \citep{macleod2010, macleod2011, macleod2012}, and those based on CRTS data using the SWV method \citep{graham2014}.  Given that we suspect the CRTS data quality to be the issue, we decided to also use the SF method with the CRTS data set to ensure mathematical framework consistent with previous studies. PS analysis would introduce a third method, and thus would be less adequate to use in our study.  

The SF for a light curve is a measure of the width of the magnitude difference distribution, as a function of the time separation, $\Delta t$ (see below for a discussion of how to account for observational errors). For two (day-averaged) epochs $j$ and $k$,  with $j > k$, the magnitude difference is computed as $\Delta m_{j,k} = m_{j} - m_{k}$, the time difference is  $\Delta t_{j,k} = t_{j} - t_{k}$, and  the combined magnitude measurement error (measurement uncertainty for $\Delta m_{j,k}$) is $e_{j,k} = (err_{j}^{2} + err_{k}^{2})^{1/2}$ (where $err_{j}$ is defined by equation~\ref{eq:2}). 

We compute SF as a function of time difference $\Delta t_{j,k}$ (hereafter, $\Delta t$ for brevity and similarly, $\Delta m$ for $\Delta m_{j,k}$ and $e$ for $e_{j,k}$) by binning ($\Delta t$, $\Delta m$, $e$) data along $\Delta t$ axis.  With a mean number of data points per light curve of  70, on average, we generate $\sum_{j=2}^{70}{(j-1)} = 2,415$  ($\Delta t$, $\Delta m$, $e$) data points. This large number allows us  to simply use 200 linearly spaced bins of $\Delta t$, which provide adequate time resolution while ensuring sufficiently large number of  $\Delta m$ values per bin.

Given that we suspect data and data processing problems as a plausible explanation for discrepant
results between SDSS-based and CRTS-based studies, we choose to study variability in the observed frame (the available SDSS
redshifts for all objects enable analysis in the rest frame, too -- see Fig.~\ref{fig:5}). 

The top two panels in Fig.~\ref{fig:2} show the standard deviation for $\Delta m$, and the robust standard 
deviation ($\sigma_G=0.741 (q_{75} - q_{25})$, where $q_{25}$ and $q_{75}$ are 25\% and 75\% quartiles) 
estimate computed from the interquartile range, as a function of $\Delta t$ for quasars, and 
separately for blue and red stars. $\sigma_G$ is somewhat smaller than the standard deviation, which indicates mild 
non-Gaussianity of $\Delta m$ distributions. For $\Delta t$ below about 100 d, all three subsamples 
show similar behaviour, while for longer time-scales quasars show appreciably larger scatter of observed 
$\Delta m$ due to intrinsic variability. In order to estimate the intrinsic variability, these ``raw'' 
measurements need to be corrected for the effects of observational (measurement) errors, as described next.

\subsection{Effects of observational errors on SF}

Given a bin with $M$ values of ($\Delta t_i$, $\Delta m_i$, $e_i$), $i=1...M$, SF will correspond to the rms width
of the $\Delta m_i$ distribution, $\sigma_{tot}$, {\it only if} all $e_i$ are negligibly small compared to the true SF 
value. When measurement uncertainties are homoscedastic, $e_i = \bar{e}$, then simply SF = $(\sigma_{tot}^2
- \bar{e}^2)^{1/2}$. In a general case of heteroscedastic uncertainties, the correction for the effects of 
observational errors is more involved because each value $\Delta m_i$ is drawn from a {\it different}
Gaussian distribution whose width is given by $\sigma_i  = (\rm{SF}^2 + e_i^2)^{1/2}$. Indeed, in this general 
case the distribution of all $\Delta m_i$ in a given bin {\it need not be a Gaussian at all!}

We refer the reader for a detailed discussion of how to estimate SF in a general case to \cite{ivezic2014}, 
and here briefly summarize the gist of their maximum likelihood method. The likelihood of a set of $M$
measurements $\Delta m_i$ is given by 
\begin{equation}
\label{eq:5}
p(\{\Delta m_{i}\} | \rm{SF}, \mu, \{e_{i}\} ) = 
 \prod _{i=1}^{M} { \frac{1}{\sqrt{2\pi} \sigma_i} \exp{\left( \frac{-(\Delta m_{i}-\mu)^{2}}{2\sigma_i^{2}} \right)}},
\end{equation}
where $\{.\}$ denotes a set of values and $\mu$ is introduced to account for possible systematic photometric 
errors between observing epochs that define the bin's $\Delta t_i$ values. We note that this expression is only 
an approximation to the true likelihood because it assumes that measurement errors for $\Delta m_i$ are 
uncorrelated. This assumption is, strictly speaking, not true because different $\Delta m_i$ values can be 
based on the same individual magnitude measurements. In practice, the covariance between errors can introduce a bias in maximum likelihood solutions, but only for $M$ much larger than used here these errors become not negligible compared to the SF. Indeed, we used the same maximum likelihood method as \cite{schmidt2010}, equation(2), that assumes no correlation between errors. 

There is no closed form solution for maximizing the likelihood given by equation~\ref{eq:5} and
we estimate SF numerically, using code\footnote{See http://www.astroml.org/book\_figures/chapter5/index.html}
from {\it astroML} python module \citep{astroML}. With the aid of Bayes Theorem and using uniform priors
for SF and $\mu$, the logarithm of the posterior probability distribution function (pdf) for SF and $\mu$ becomes
\begin{equation}
\label{eq:6}
L_p(SF,\mu) = {\rm const.} - {1\over2} \sum_{i=1}^M \left( \ln(\rm{SF}^2+e_i^2)
               + {  (\Delta m_i-\mu)^2 \over \rm{SF}^2+e_i^2} \right).
\end{equation}
We evaluate $L_{p}$ on a grid\footnote{The grid size is set using approximate solutions described by \cite{ivezic2014}.}
of $\mu$ and SF first, find its maximum that yields the maximum a posteriori (MAP) estimates for SF and $\mu$, 
and then marginalize over $\mu$ to find the posterior pdf for SF as 
\begin{equation}
p(\rm{SF}) = \int_{0}^{\infty} {p(\rm{SF}, \mu |\{\Delta m_{i}\}, \{e_{i}\})} d \mu,
\end{equation}
which is used to estimate the uncertainty (the credible region) of MAP estimate for SF. When there is no strong 
evidence for intrinsic variability, SF tends to zero.

\begin{figure}
\includegraphics[width=1.05\columnwidth, center]{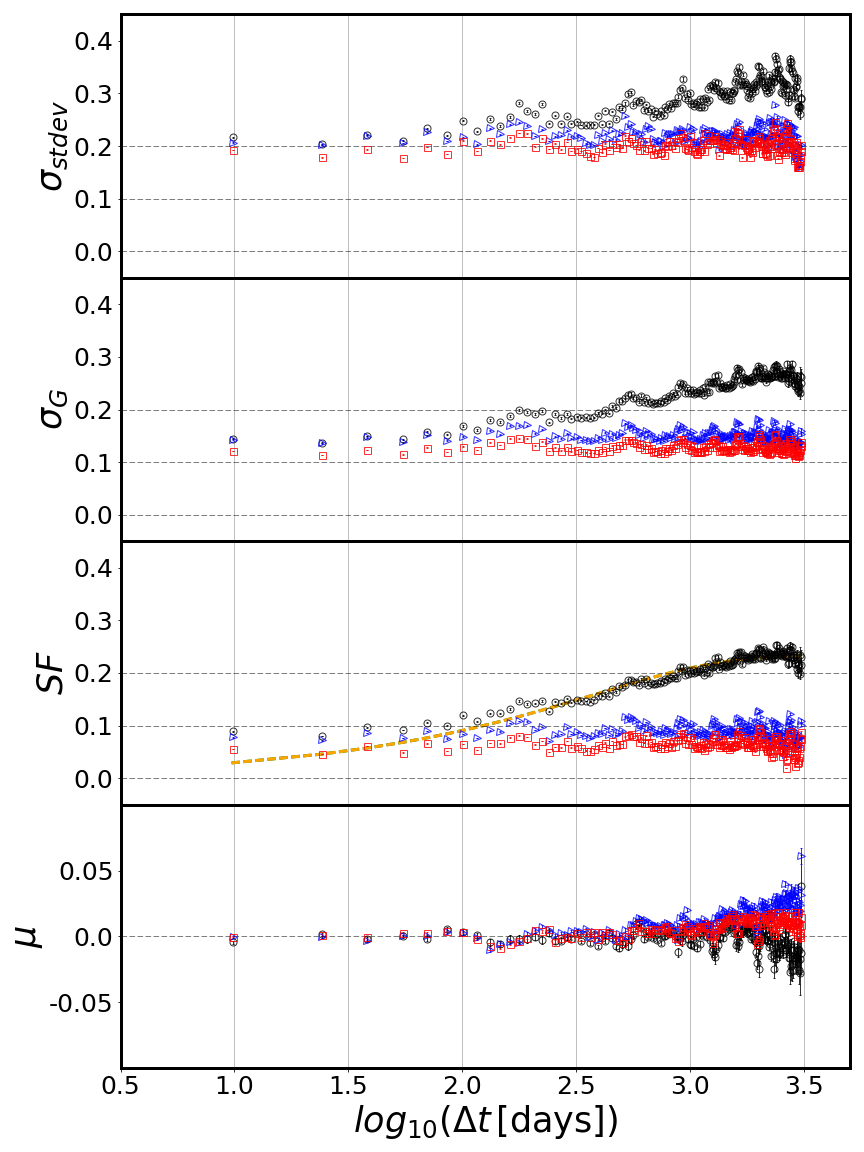}
\vskip -0.15in
\caption{The four panels show various statistics computed for subsamples of 747 CRTS quasars (black circles), 
1496 `blue' stars (blue triangles), and 2327 `red' stars (red squares), with SDSS $r$ magnitudes in the range 18.5-19.
 Red and blue stars have SDSS colours $1 < \mathrm{g-i} < 3$ and  $-1 < \mathrm{g-i} < 1$, respectively. All pairwise CRTS brightness 
differences are binned in 200 linearly spaced bins of time difference $\Delta t$. For each bin, we compute,
from top to bottom: the standard deviation $\sigma_{stdev}$, the robust standard deviation estimate $\sigma_{G}$ based on 
the interquartile range , the structure function SF, and the mean value of $\Delta m$ per bin $\mu$. The statistical
(random) errors are often smaller than the symbol size due to large number of data points; systematic errors for
all displayed quantities are probably of the order 0.01 mag (not shown). Both $\mu$ and SF are found from the 
two-dimensional maximum of the log-likelihood $L_{p}$ on the $[\mu,\rm{SF}]$ grid (see equation~\ref{eq:6}).  
The yellow dashed line in the third panel traces the fiducial DRW model (see equation~\ref{eq:DRWSF}). We address the peculiar wiggle behaviour in the Appendix~\ref{sec:wiggles}, but it does not have any influence on our overall conclusions. }
\label{fig:2}
\end{figure}

The bottom two panels in Fig.~\ref{fig:2} show SF and $\mu$ as a function of $\Delta t$ for quasars, blue 
and red stars. For $\Delta t$ below about 1000 d, $\mu$ for all three subsamples is within 0.01 mag
from zero, as expected. On the other hand, SF below about 100 d is in the range 0.05-0.10 mag for
all three subsamples. In the case of quasars, the observed SF$\sim$0.1 mag for \mbox{10 d $<\Delta t<$100 d}
demonstrates that the difference between SDSS results from \cite{macleod2010} (see the yellow dashed
line in the third panel) and CRTS results from \cite{graham2014} is {\it not} due to different analysis 
methods (SF versus SWV, respectively): {\it Here, we fully reproduce this discrepancy using the SF method and 
CRTS data.} 

Fig.~\ref{fig:2} also indicates a plausible solution to this puzzle: the observed SF for both blue and red
stars in the range \mbox{10$^d<\Delta t<$100$^d$} is {\it unexpectedly} large: The values are in the range 
0.05--0.10 mag rather than negligible (say, $\la 0.01-0.02$ mag). In other words, more variation is 
observed in light curves of non-variable stars than can be explained with reported photometric errors.
The same result is obtained for all three chosen magnitude bins. Such a behaviour could be observed if 
photometric error estimates computed by the CRTS photometric pipeline are mis-estimated, resulting
in an incorrect correction for observational errors. We proceed to perform an independent test of 
photometric errors using repeated observations of non-variable standard stars.

\subsection{Tests of observational errors using non-variable stars}
\label{sec:results}

\begin{figure}
\includegraphics[width=1.1\columnwidth, center]{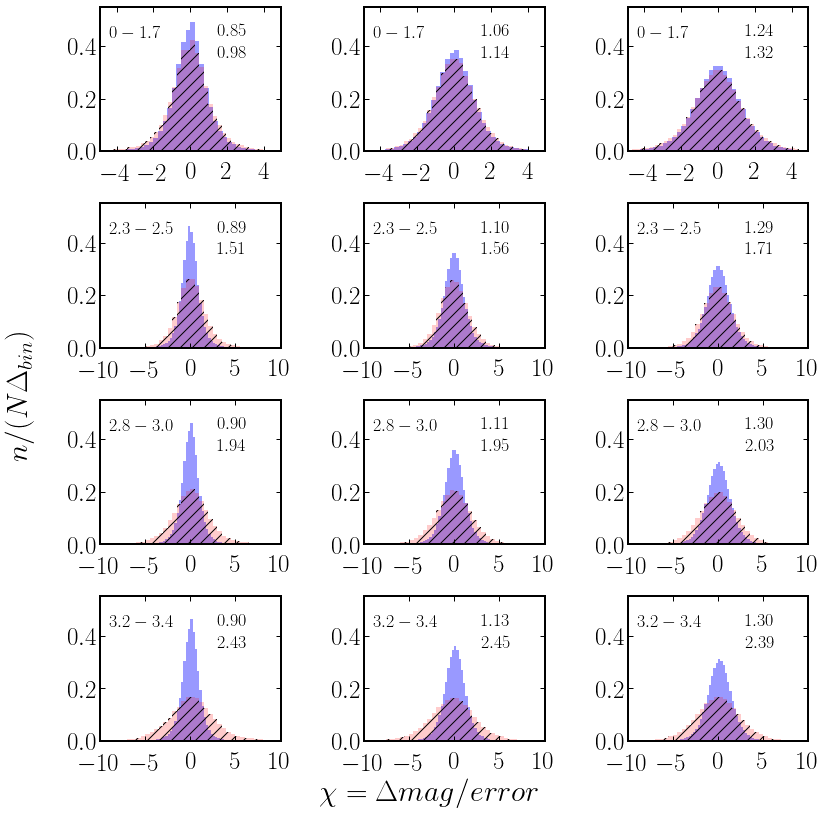}
\vskip -0.15in
\caption{Histograms show CRTS-based $\chi = \Delta m / \mathrm{error}$ for blue stars (blue shading) 
and quasars (red hatched shading), split into bins of $\log{\Delta t}$ (rows) and SDSS $r$ magnitude (columns). 
Vertically, from top to bottom, $\log{\Delta t}$ : $0<\log{\Delta t}<1.7$ ($t < 50 $ days), 
$2.3<\log{\Delta t}<2.5$, $2.8<\log{\Delta t}<3.0$, and $3.2<\log{\Delta t}<3.4$ (indicated by numbers
 in the upper left-hand corner of each subplot). Horizontally, from the left- to right-hand side, the SDSS $r$ magnitude bins are:  $17-18$,  $18-18.5$, and $18.5-19$. The numbers in the upper right-hand corner of each subplot are the 
robust width of $\chi$ distributions determined using interquartile range ($\sigma_G$); upper value for
blue stars and lower value for quasars.}
\label{fig:3}
\end{figure}

Assuming that standard stars from SDSS are truly non-variable, if (Gaussian) photometric error estimates 
computed by the CRTS photometric pipeline are correct, then the distribution of $\chi_i= \Delta m_i / e_i$ 
for stars should be distributed as a unit Gaussian, $N$(0,1). Deviations of the distribution width for stars 
from unity indicate incorrect photometric error estimates. For quasars, we expect that the width should
exceed unity because of their intrinsic variability, and that the width should increase with $\Delta t$. 
We perform this test in Fig.~\ref{fig:3}, where we show $\chi$ distributions for both blue stars and quasars, 
and for a grid of $\Delta t$ and magnitude bins. 

For the shortest $\Delta t$ bin ($<$50 d), the distributions for stars and quasars appear indistinguishable
for all three magnitude bins. This similarity immediately argues that there is no detected intrinsic variability
for quasars. Furthermore, the width of $\chi$ distributions for stars appears to be a function of magnitude,
with very little dependence on $\Delta t$. The distribution widths for stars in each magnitude bin (all $\Delta t$
values), obtained using robust width estimator $\sigma_G$, are listed in Table~\ref{tab:fc}. For example,
the bin with $18.5<r<19$, which contains the majority of quasars, appears to have under-estimated photometric
errors by a factor of 1.3, on average. The same conclusion is derived using red stars. For small $\Delta t$, where 
quasar SF is intrinsically small, the quasar SF will be thus significantly over-estimated, while for large $\Delta t$, 
where the quasar SF is intrinsically large, the effect on SF will be small. We extend this qualitative conclusion 
to a more quantitative analysis in the next section. 

We note that problems with CRTS photometric uncertainty estimates have been reported before
(e.g., \citealt{vaughan2016}). Additional analysis of CRTS photometric uncertainty estimates, beyond magnitude
limits of direct interest to quasar variability analysis, is presented in Appendix~\ref{sec:crts_photometry}. 

\begin{table}
\centering
\begin{threeparttable}
\caption{The robust distribution widths for $\chi$ for blue stars.}
\label{tab:fc}
\begin{tabular}{ lr } 
\hline
 Magnitude  $\,\,\,\,\,\,\,\,\,\,\,\,\,\,\,\,\,\,\,\,$ & $\sigma_G$ \\ 
 \hline
17--18   & 0.870   \\
18--18.5 & 1.107   \\
18.5--19  & 1.288   \\
\hline
\end{tabular}
\end{threeparttable}
\end{table}

\subsection{SF with corrected observational errors}

Informed by the analysis from preceding section, we assume that correction factors for photometric
error estimates are independent of colour and are only a function of magnitude. Depending on the magnitude
of stars and quasars, we multiply their reported photometric errors by $\sigma_G$ values listed in 
Table~\ref{tab:fc}, and repeat SF analysis. By construction, we expect that the width of $\chi$ distributions
for blue stars will be unity, and that their SF will tend to 0. For quasars, compared to SF values shown in the third 
panel in Fig.~\ref{fig:2}, we expect somewhat smaller SF at large $\Delta t$ and much smaller SF at small $\Delta t$. 

\begin{figure}
\vskip -0.15in
\includegraphics[width=1.1\columnwidth, center]{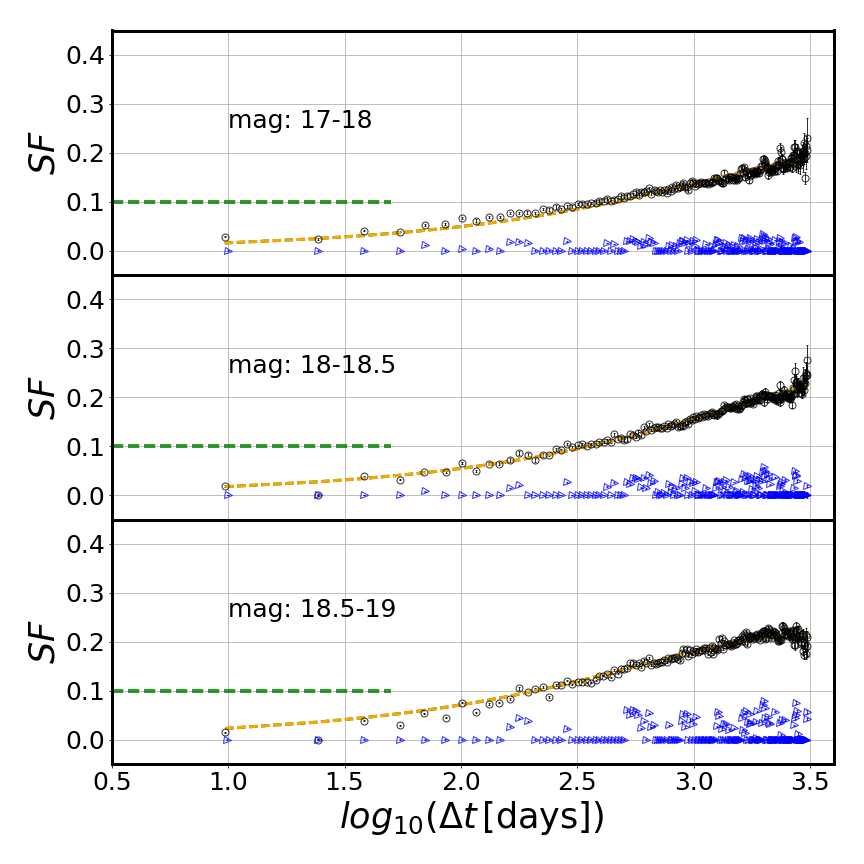}
\caption{Analogous to the third panel in Fig.~\ref{fig:2}, except that here SF for blue stars and quasars in
all three magnitude bins are shown, and photometric errors are modified by multiplicative correction factors 
listed in Table~\ref{tab:fc}. Note that SF for stars in vanishing, while SF for quasars at $\log_{10}(\Delta t) < 1.7$  
is about twice as small as in Fig.~\ref{fig:2}.} 
\label{fig:4}
\end{figure}

\begin{figure}

\vskip -0.15in
\includegraphics[width=1.1\columnwidth, center]{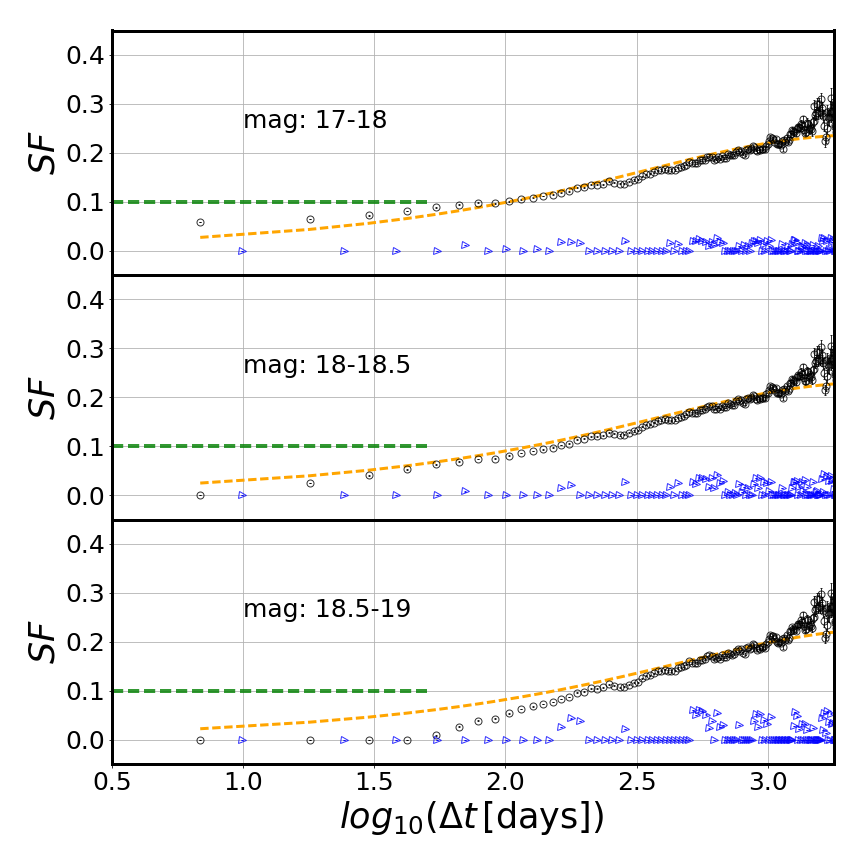}
\caption{Analogous to Fig.~\ref{fig:4}, but here for $\Delta t$ in the quasar rest frame : t\textsubscript{rest} = t\textsubscript{obs} / (1+$z$), using known quasar redshifts from SDSS \citep{macleod2010}.  The rest frame correction shifts time lags to shorter time-scales and produces SF for quasars in agreement with corresponding results obtained by \citealt{macleod2010}.}
\label{fig:5}
\end{figure}

Fig.~\ref{fig:4} shows SF for blue stars and quasars for subsamples from the three selected magnitude bins. 
As evident, both expectations are born out: for all three magnitude bins, SF for blue stars is essentially 
vanishing within noise ($\sim$0.05 mag), while SF for quasars at small $\Delta t$ is about twice smaller 
than in Fig.~\ref{fig:2} and thus consistent with the values based on SDSS data. In Fig.~\ref{fig:5}, we 
demonstrate that this agreement with SDSS results extends to rest frame analysis, too. 

\subsection{SF estimated from PTF data}

Recent PTF (Palomar Transient Factory) Data Release 3 light curves\footnote{\url{http://www.ptf.caltech.edu/page/lcdb} \citep{rau2009}}
can be used for an independent test of our conclusions derived above. We queried the NASA/IPAC Infrared Science 
Archive\footnote{\url{https://irsa.ipac.caltech.edu}} `PTF Objects' catalogue using coordinates for 7601 spectroscopically
confirmed Stripe 82 quasars, and 48$\,$250 standard stars (same as the final samples used for CRTS-based analysis).
A positional multi-object search with a matching radius of 2 arcsec, with a flag `ngoodobs' $>$ 10,  resulted in  
6471 quasars and 38$\,$776 stars.  For these objects, we obtained time series data from the `PTF Light Curve Table' catalogue
(we grouped by SDSS coordinates). 

We processed these PTF light curves in exactly the same way as the CRTS light curves. We first performed day-averaging, 
using the weighted error as the measure of uncertainty on day-averaged brightness measurement. We further selected only 
those objects that have been observed on at least 10 different nights, resulting in samples of 2753 quasars and 15$\,$714 stars. 
The counts of magnitude-limited subsamples are listed in Table~\ref{tab:ptf}. 

\begin{table}
\centering
\begin{threeparttable}
\caption{Count of stars and quasars, selected by their SDSS $r$ magnitudes and $g$--$i$ colours. Analoguous to Table~\ref{tab:object_count}, except that here the counts of stars and quasars with PTF
adequate data are listed.}
\label{tab:ptf}
\begin{tabular}{ l ccc } 
\hline
$r$ magnitude  & Red stars & Blue stars & Quasars \\ 
\hline
17--18   & 1243 & 1077   & 90    \\ 
18--18.5 & 825 &  497  & 160   \\ 
18.5--19 & 913 &  548  & 377   \\
 & & & \\
Total       & 2981 &  2122 & 627  \\
\hline
\end{tabular}
\end{threeparttable}
\end{table}
 
\begin{figure}
\includegraphics[width=1.1\columnwidth, center]{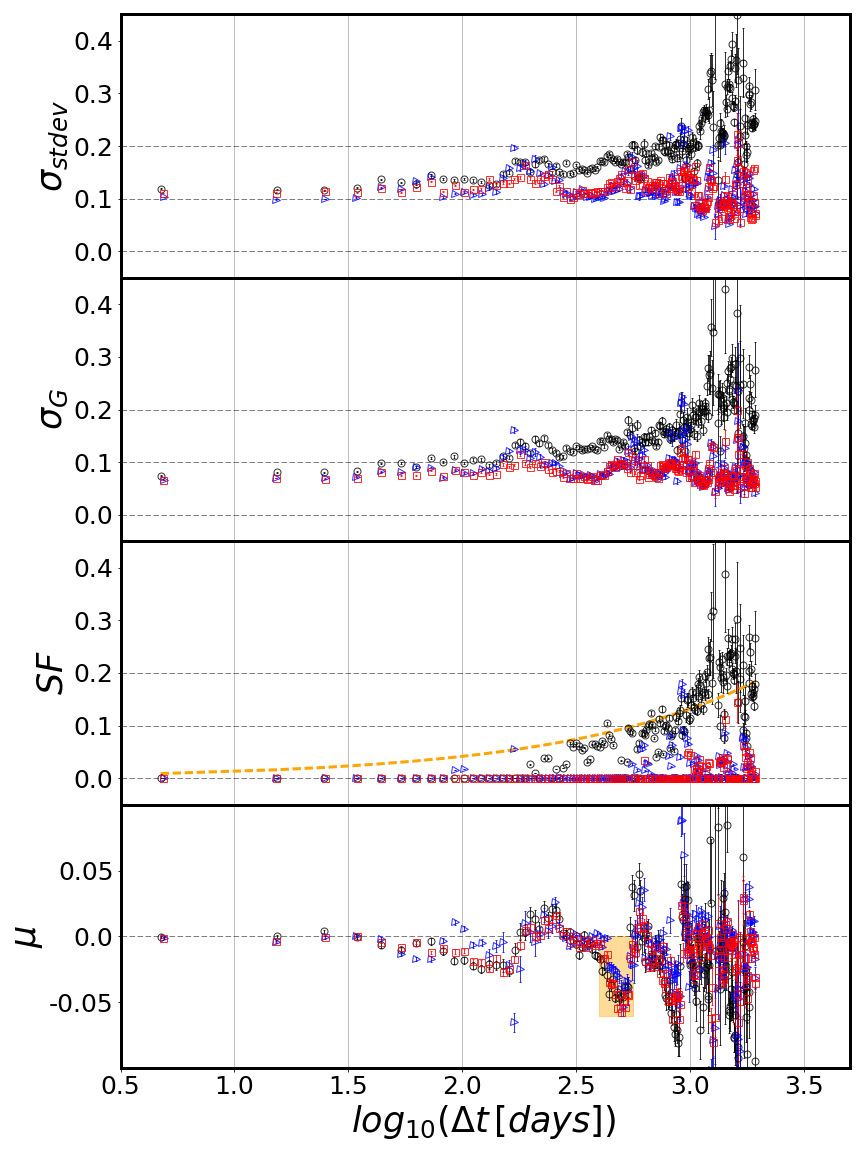}
\caption{Analogous to Fig.~\ref{fig:2}, but here the statistics for subsamples of 377 quasars (black circles), 548 `blue' stars (blue triangles), and 913 `red' stars (red squares), with adequate PTF light curve data are shown. Note that the mean magnitude difference ($\mu$, the bottom panel) does not stay as close to 0 as for CRTS data -- a deviation around $\log_{10} \Delta t \approx 2.7$ might indicate some issues with photometric zero-point calibration (at the level of 0.02--0.03 mag).}
\label{fig:2PTF} 
\end{figure}

The SF results based on PTF light curve data are shown in Fig.~\ref{fig:2PTF}.  For these uncorrected PTF data, 
it is evident that there is no sign of variability for quasars on short time-scales ($\Delta t < 100 \, d$) 
above the SDSS-level of $\sim$0.05 mag (unlike for CRTS data, see Fig.~\ref{fig:2}). Note also that standard stars 
show no appreciable variability at any time-scale (SF$\approx0$). Therefore, this PTF-based analysis further 
supports our conclusion that extraneous quasar variability at short time-scales was due to slightly underestimated
photometric uncertainties. 

\section{Conclusions}

We analysed the error properties of the CRTS sample of quasars and standard stars. Using 
repeated CRTS observations of non-variable stars, we found that the photometric error estimates 
computed by the CRTS photometric pipeline are slightly under-estimated for the majority of
quasars. When quasar light curves are corrected for the impact of observational errors, the
resulting corrections to the SF are thus too small. For small $\Delta t$, where quasar SF is intrinsically 
small, quasar SF is significantly over-estimated (akin to the subtraction of two large numbers 
to get a small number, when the second large number is under-estimated). In particular, at time-scales of about 50 d, SF is over-estimated by about a factor of 2. This behaviour provides 
a plausible explanation for the increased quasar variability level in CRTS light curves reported 
by \cite{graham2014}, compared to earlier SDSS-based results obtained by \cite{macleod2010}.
An additional analysis based on independent light curve data for the same objects obtained by 
the PTF provides further support for this conclusion. We conclude that
the quasar variability constraints on weekly and monthly time-scales from SDSS, CRTS and PTF 
surveys are mutually compatible, as well as consistent with DRW model.

\section*{Acknowledgements}

We thank Eric Bellm for his help with the PTF data retrieval and reduction of light curves.  
We thank Neven Caplar for fruitful discussions about the use of PTF data and SF methodology. 

Funding for the SDSS and SDSS-II has been provided by the Alfred P. Sloan Foundation, 
the Participating Institutions, the National Science Foundation, the U.S. Department of 
Energy, the National Aeronautics and Space Administration, the Japanese Monbukagakusho,
the Max Planck Society, and the Higher Education Funding Council for England. The SDSS 
Web Site is http://www.sdss.org/.

The SDSS is managed by the Astrophysical Research Consortium for the Participating Institutions. 
The Participating Institutions are the American Museum of Natural History, Astrophysical Institute 
Potsdam, University of Basel, University of Cambridge, Case Western Reserve University, University
of Chicago, Drexel University, Fermilab, the Institute for Advanced Study, the Japan Participation 
Group, Johns Hopkins University, the Joint Institute for Nuclear Astrophysics, the Kavli Institute for 
Particle Astrophysics and Cosmology, the Korean Scientist Group, the Chinese Academy of Sciences 
(LAMOST), Los Alamos National Laboratory, the Max-Planck-Institute for Astronomy (MPIA), the 
Max-Planck-Institute for Astrophysics (MPA), New Mexico State University, Ohio State University, 
University of Pittsburgh, University of Portsmouth, Princeton University, the United States Naval
Observatory, and the University of Washington.

\bibliographystyle{mnras}
\bibliography{references} 

\begin{thebibliography}{}
\makeatletter
\relax
\def\mn@urlcharsother{\let\do\@makeother \do\$\do\&\do\#\do\^\do\_\do\%\do\~}
\def\mn@doi{\begingroup\mn@urlcharsother \@ifnextchar [ {\mn@doi@}
  {\mn@doi@[]}}
\def\mn@doi@[#1]#2{\def\@tempa{#1}\ifx\@tempa\@empty \href
  {http://dx.doi.org/#2} {doi:#2}\else \href {http://dx.doi.org/#2} {#1}\fi
  \endgroup}
\def\mn@eprint#1#2{\mn@eprint@#1:#2::\@nil}
\def\mn@eprint@arXiv#1{\href {http://arxiv.org/abs/#1} {{\tt arXiv:#1}}}
\def\mn@eprint@dblp#1{\href {http://dblp.uni-trier.de/rec/bibtex/#1.xml}
  {dblp:#1}}
\def\mn@eprint@#1:#2:#3:#4\@nil{\def\@tempa {#1}\def\@tempb {#2}\def\@tempc
  {#3}\ifx \@tempc \@empty \let \@tempc \@tempb \let \@tempb \@tempa \fi \ifx
  \@tempb \@empty \def\@tempb {arXiv}\fi \@ifundefined
  {mn@eprint@\@tempb}{\@tempb:\@tempc}{\expandafter \expandafter \csname
  mn@eprint@\@tempb\endcsname \expandafter{\@tempc}}}

\bibitem[\protect\citeauthoryear{{Abazajian} et~al.,}{{Abazajian}
  et~al.}{2009}]{abazajian2009}
{Abazajian} K.~N.,  et~al., 2009, \mn@doi [\apjs]
  {10.1088/0067-0049/182/2/543}, \href
  {http://adsabs.harvard.edu/abs/2009ApJS..182..543A} {182, 543}

\bibitem[\protect\citeauthoryear{{Charisi}, {Bartos}, {Haiman}, {Price-Whelan},
  {Graham}, {Bellm}, {Laher}  \& {M{\'a}rka}}{{Charisi}
  et~al.}{2016}]{charisi2016}
{Charisi} M.,  {Bartos} I.,  {Haiman} Z.,  {Price-Whelan} A.~M.,  {Graham}
  M.~J.,  {Bellm} E.~C.,  {Laher} R.~R.,   {M{\'a}rka} S.,  2016, \mn@doi
  [\mnras] {10.1093/mnras/stw1838}, \href
  {http://adsabs.harvard.edu/abs/2016MNRAS.463.2145C} {463, 2145}

\bibitem[\protect\citeauthoryear{{Collier} \& {Peterson}}{{Collier} \&
  {Peterson}}{2001}]{peterson2001}
{Collier} S.,  {Peterson} B.~M.,  2001, \mn@doi [\apj] {10.1086/321517}, \href
  {http://adsabs.harvard.edu/abs/2001ApJ...555..775C} {555, 775}

\bibitem[\protect\citeauthoryear{{Djorgovski} et~al.,}{{Djorgovski}
  et~al.}{2011}]{djorgovski2011a}
{Djorgovski} S.~G.,  et~al., 2011, preprint (arXiv:1102.5004), \href
  {http://adsabs.harvard.edu/abs/2011arXiv1102.5004D} {}

\bibitem[\protect\citeauthoryear{{Drake} et~al.,}{{Drake}
  et~al.}{2013}]{drake2013}
{Drake} A.~J.,  et~al., 2013, \mn@doi [\apj] {10.1088/0004-637X/763/1/32},
  \href {http://adsabs.harvard.edu/abs/2013ApJ...763...32D} {763, 32}

\bibitem[\protect\citeauthoryear{{Graham}, {Drake}, {Djorgovski}, {Mahabal},
  {Donalek}, {Duan}  \& {Maker}}{{Graham} et~al.}{2013}]{graham2013}
{Graham} M.~J.,  {Drake} A.~J.,  {Djorgovski} S.~G.,  {Mahabal} A.~A.,
  {Donalek} C.,  {Duan} V.,   {Maker} A.,  2013, \mn@doi [\mnras]
  {10.1093/mnras/stt1264}, \href
  {http://adsabs.harvard.edu/abs/2013MNRAS.434.3423G} {434, 3423}

\bibitem[\protect\citeauthoryear{{Graham}, {Djorgovski}, {Drake}, {Mahabal},
  {Chang}, {Stern}, {Donalek}  \& {Glikman}}{{Graham}
  et~al.}{2014}]{graham2014}
{Graham} M.~J.,  {Djorgovski} S.~G.,  {Drake} A.~J.,  {Mahabal} A.~A.,  {Chang}
  M.,  {Stern} D.,  {Donalek} C.,   {Glikman} E.,  2014, \mn@doi [\mnras]
  {10.1093/mnras/stt2499}, \href
  {http://adsabs.harvard.edu/abs/2014MNRAS.439..703G} {439, 703}

\bibitem[\protect\citeauthoryear{{Graham} et~al.,}{{Graham}
  et~al.}{2015}]{graham2015b}
{Graham} M.~J.,  et~al., 2015, \mn@doi [\mnras] {10.1093/mnras/stv1726}, \href
  {http://adsabs.harvard.edu/abs/2015MNRAS.453.1562G} {453, 1562}

\bibitem[\protect\citeauthoryear{{Hawkins}}{{Hawkins}}{2002}]{hawkins2002}
{Hawkins} M.~R.~S.,  2002, \mn@doi [\mnras] {10.1046/j.1365-8711.2002.04939.x},
  \href {http://adsabs.harvard.edu/abs/2002MNRAS.329...76H} {329, 76}

\bibitem[\protect\citeauthoryear{{Hawkins}}{{Hawkins}}{2007}]{hawkins2007}
{Hawkins} M.~R.~S.,  2007, \mn@doi [\aap] {10.1051/0004-6361:20066283}, \href
  {http://adsabs.harvard.edu/abs/2007A%26A...462..581H} {462, 581}

\bibitem[\protect\citeauthoryear{{Ivezi\'c} et~al.,}{{Ivezi\'c}
  et~al.}{2004}]{2004Ivezic}
{Ivezi\'c} {\v Z}.,  et~al., 2004, in {Storchi-Bergmann} T.,  {Ho} L.~C.,
  {Schmitt} H.~R.,  eds,  IAU Symposium Vol. 222, The Interplay Among Black
  Holes, Stars and ISM in Galactic Nuclei. pp 525--526 (\mn@eprint {}
  {astro-ph/0404487}), \mn@doi{10.1017/S1743921304003126}

\bibitem[\protect\citeauthoryear{{Ivezi{\'c}} et~al.,}{{Ivezi{\'c}}
  et~al.}{2007}]{ivezic2007}
{Ivezi{\'c}} {\v Z}.,  et~al., 2007, \mn@doi [\aj] {10.1086/519976}, \href
  {http://adsabs.harvard.edu/abs/2007AJ....134..973I} {134, 973}

\bibitem[\protect\citeauthoryear{{Ivezi{\'c}}, {Connolly}, {VanderPlas}  \&
  {Gray}}{{Ivezi{\'c}} et~al.}{2014}]{ivezic2014}
{Ivezi{\'c}} {\v Z}.,  {Connolly} A.~J.,  {VanderPlas} J.~T.,   {Gray} A.,
  2014, {Statistics, Data Mining, and Machine Learning in Astronomy, Princeton
  Univ. Press, Princeton and Oxford}

\bibitem[\protect\citeauthoryear{{Kawaguchi}, {Mineshige}, {Umemura}  \&
  {Turner}}{{Kawaguchi} et~al.}{1998}]{kawaguchi1998}
{Kawaguchi} T.,  {Mineshige} S.,  {Umemura} M.,   {Turner} E.~L.,  1998,
  \mn@doi [\apj] {10.1086/306105}, \href
  {http://adsabs.harvard.edu/abs/1998ApJ...504..671K} {504, 671}

\bibitem[\protect\citeauthoryear{{Kelly}, {Bechtold}, {Siemiginowska},
  {Aldcroft}  \& {Sobolewska}}{{Kelly} et~al.}{2007}]{kelly2007}
{Kelly} B.~C.,  {Bechtold} J.,  {Siemiginowska} A.,  {Aldcroft} T.,
  {Sobolewska} M.,  2007, \mn@doi [\apj] {10.1086/510876}, \href
  {http://adsabs.harvard.edu/abs/2007ApJ...657..116K} {657, 116}

\bibitem[\protect\citeauthoryear{Kelly, Bechtold  \& Siemiginowska}{Kelly
  et~al.}{2009}]{kelly2009}
Kelly B.~C.,  Bechtold J.,   Siemiginowska A.,  2009, The Astrophysical
  Journal, 698, 895

\bibitem[\protect\citeauthoryear{{Kelly}, {Sobolewska}  \&
  {Siemiginowska}}{{Kelly} et~al.}{2011}]{kelly2011}
{Kelly} B.~C.,  {Sobolewska} M.,   {Siemiginowska} A.,  2011, \mn@doi [\apj]
  {10.1088/0004-637X/730/1/52}, \href
  {http://adsabs.harvard.edu/abs/2011ApJ...730...52K} {730, 52}

\bibitem[\protect\citeauthoryear{{Koz{\l}owski}}{{Koz{\l}owski}}{2016}]{kozlowski2016}
{Koz{\l}owski} S.,  2016, \mn@doi [\apj] {10.3847/0004-637X/826/2/118}, \href
  {http://adsabs.harvard.edu/abs/2016ApJ...826..118K} {826, 118}

\bibitem[\protect\citeauthoryear{{Lawrence}}{{Lawrence}}{2016}]{lawrence2016a}
{Lawrence} A.,  2016, in {Mickaelian} A.,  {Lawrence} A.,   {Magakian} T.,
  eds,  Astronomical Society of the Pacific Conference Series Vol. 505,
  Astronomical Surveys and Big Data. p.~107 (\mn@eprint {arXiv} {1605.09331})

\bibitem[\protect\citeauthoryear{MacLeod et~al.,}{MacLeod
  et~al.}{2010}]{macleod2010}
MacLeod C.~L.,  et~al., 2010, The Astrophysical Journal, 721, 1014

\bibitem[\protect\citeauthoryear{MacLeod et~al.,}{MacLeod
  et~al.}{2011}]{macleod2011}
MacLeod C.~L.,  et~al., 2011, The Astrophysical Journal, 728, 26

\bibitem[\protect\citeauthoryear{MacLeod et~al.,}{MacLeod
  et~al.}{2012}]{macleod2012}
MacLeod C.~L.,  et~al., 2012, The Astrophysical Journal, 753, 106

\bibitem[\protect\citeauthoryear{{Rau} et~al.,}{{Rau} et~al.}{2009}]{rau2009}
{Rau} A.,  et~al., 2009, \mn@doi [\pasp] {10.1086/605911}, \href
  {http://adsabs.harvard.edu/abs/2009PASP..121.1334R} {121, 1334}

\bibitem[\protect\citeauthoryear{{Schmidt}, {Marshall}, {Rix}, {Jester},
  {Hennawi}  \& {Dobler}}{{Schmidt} et~al.}{2010}]{schmidt2010}
{Schmidt} K.~B.,  {Marshall} P.~J.,  {Rix} H.-W.,  {Jester} S.,  {Hennawi}
  J.~F.,   {Dobler} G.,  2010, \mn@doi [\apj] {10.1088/0004-637X/714/2/1194},
  \href {http://adsabs.harvard.edu/abs/2010ApJ...714.1194S} {714, 1194}

\bibitem[\protect\citeauthoryear{{Schneider} et~al.,}{{Schneider}
  et~al.}{2010}]{schneider2010}
{Schneider} D.~P.,  et~al., 2010, VizieR Online Data Catalog, \href
  {http://adsabs.harvard.edu/abs/2010yCat.7260....0S} {7260}

\bibitem[\protect\citeauthoryear{{Vanden Berk} et~al.,}{{Vanden Berk}
  et~al.}{2004}]{berk2004}
{Vanden Berk} D.~E.,  et~al., 2004, \mn@doi [\apj] {10.1086/380563}, \href
  {http://adsabs.harvard.edu/abs/2004ApJ...601..692V} {601, 692}

\bibitem[\protect\citeauthoryear{{Vanderplas}, {Connolly}, {Ivezi{\'c}}  \&
  {Gray}}{{Vanderplas} et~al.}{2012}]{astroML}
{Vanderplas} J.,  {Connolly} A.,  {Ivezi{\'c}} {\v Z}.,   {Gray} A.,  2012, in
  Conference on Intelligent Data Understanding (CIDU). pp 47 --54,
  \mn@doi{10.1109/CIDU.2012.6382200}

\bibitem[\protect\citeauthoryear{{Vaughan}, {Uttley}, {Markowitz},
  {Huppenkothen}, {Middleton}, {Alston}, {Scargle}  \& {Farr}}{{Vaughan}
  et~al.}{2016}]{vaughan2016}
{Vaughan} S.,  {Uttley} P.,  {Markowitz} A.~G.,  {Huppenkothen} D.,
  {Middleton} M.~J.,  {Alston} W.~N.,  {Scargle} J.~D.,   {Farr} W.~M.,  2016,
  \mn@doi [\mnras] {10.1093/mnras/stw1412}, \href
  {http://adsabs.harvard.edu/abs/2016MNRAS.461.3145V} {461, 3145}

\bibitem[\protect\citeauthoryear{{Zu}, {Kochanek}, {Koz{\l}owski}  \&
  {Udalski}}{{Zu} et~al.}{2013}]{zu2013}
{Zu} Y.,  {Kochanek} C.~S.,  {Koz{\l}owski} S.,   {Udalski} A.,  2013, \mn@doi
  [\apj] {10.1088/0004-637X/765/2/106}, \href
  {http://adsabs.harvard.edu/abs/2013ApJ...765..106Z} {765, 106}

\bibitem[\protect\citeauthoryear{{de Vries}, {Becker}, {White}  \&
  {Loomis}}{{de Vries} et~al.}{2005}]{devries2005}
{de Vries} W.~H.,  {Becker} R.~H.,  {White} R.~L.,   {Loomis} C.,  2005,
  \mn@doi [\aj] {10.1086/427393}, \href
  {http://adsabs.harvard.edu/abs/2005AJ....129..615D} {129, 615}

\makeatother
\end{thebibliography}

\appendix
\section{Variation of the CRTS photometric uncertainty with magnitude} 
\label{sec:crts_photometry}

We found in Section~\ref{sec:results} (see Table~2) that reported CRTS photometric uncertainty estimates 
are too large by $\sim$15\% in the magnitude range 17--18, and too small by $\sim$10-25\% in the 
magnitude range 18--19. Such problems have been reported before; for example, \cite{vaughan2016}
reported that for bright objects (magnitude $\sim$15) the error bars provided by the CRTS pipeline 
processing are overestimated by a factor of 4-5. Since this factor is much larger than we obtained
for fainter magnitude bins, we extend our standard star analysis to the full CRTS magnitude range. 

\begin{figure}
\includegraphics[width=1.0\columnwidth, center]{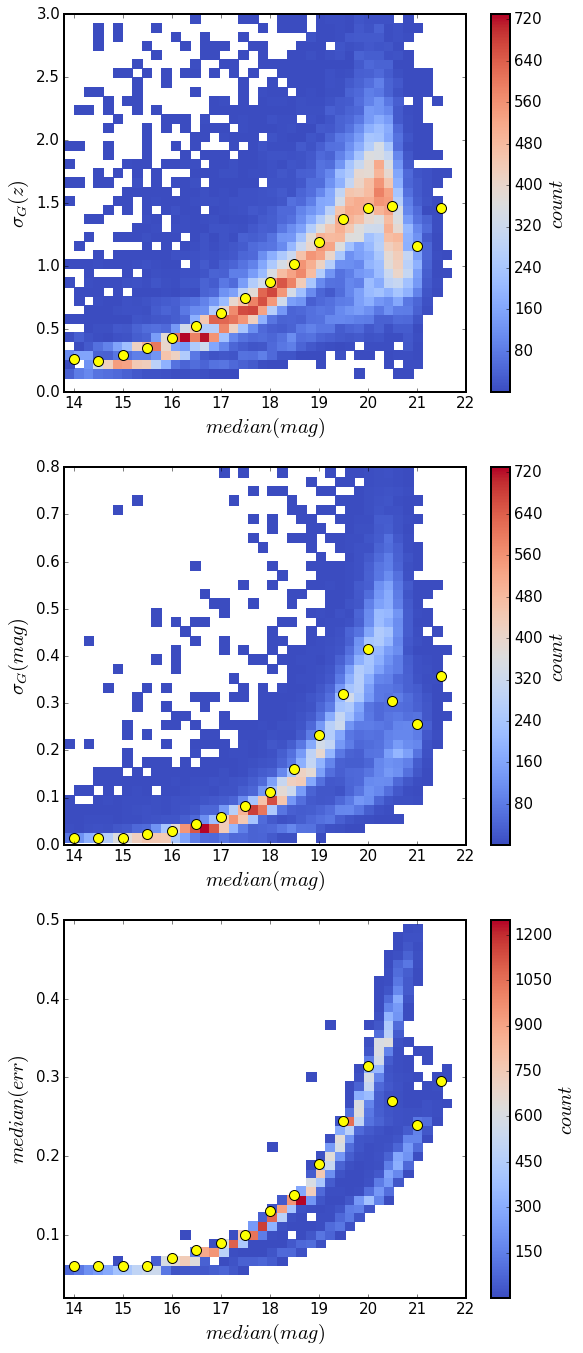}
\caption{The top panel shows the variation with magnitude of the photometric
scatter per light curve, normalized by reported CRTS photometric uncertainties 
(see equation~\ref{eq:z} for definition), using CRTS light curves for $\sim$48,000 standard 
(non-variable) stars from the SDSS catalogue. If the reported CRTS photometric uncertainties
were correctly estimated, the $\sigma_G(z)$ distribution would be centred on unity and 
independent of magnitude. The middle panel shows the observed intrinsic 
scatter per light curve, and the bottom panel shows the distribution of reported photometric 
uncertainty, both as function of median magnitude (per light curve). 
\label{fig:CRTSerrors}} 
\end{figure}

The top panel in Fig.~\ref{fig:CRTSerrors} shows the variation with magnitude of the robust distribution 
width for the quantity 
\begin{equation} 
\label{eq:z} 
           z_{ij} = { m_{ij} - m_j \over err_{ij} },
\end{equation}
where  $m_{j}$ is the weighted mean magnitude for star indexed $j$, and index $i$ runs over all
observations of a given star. The quantity $\sigma_G(z)_j$ is the robust quartile-based distribution 
width of $z_{ij}$ for a given star $j$. If the reported CRTS photometric uncertainties ($err_{ij}$) were 
correctly estimated, the $\sigma_G(z)$ distribution for standard (non-variable) stars would be 
centred on unity and independent of magnitude. As the  top panel in Fig.~\ref{fig:CRTSerrors} 
clearly demonstrates this is not the case: $\sigma_G(z)$ is $\sim$0.25 at the bright end, and 
increases to $\sim$1.5 at the faint end. In the magnitude range 17--19, the $\sigma_G(z)$ 
behaviour is consistent with the results listed in Table~2. 

The middle and bottom panels show that the observed intrinsic scatter per light curve at the bright
end is $\sim0.01$ mag, while reported photometric uncertainty is never smaller than 0.05 mag. 
In other words, we confirm the result reported by \cite{vaughan2016} for the bright end and
demonstrate that problems with reported CRTS photometric uncertainties are a strong function
of magnitude. 

\section{CSS calibration wiggles} 
\label{sec:wiggles}

\begin{figure}
\includegraphics[width=1.0\columnwidth, center]{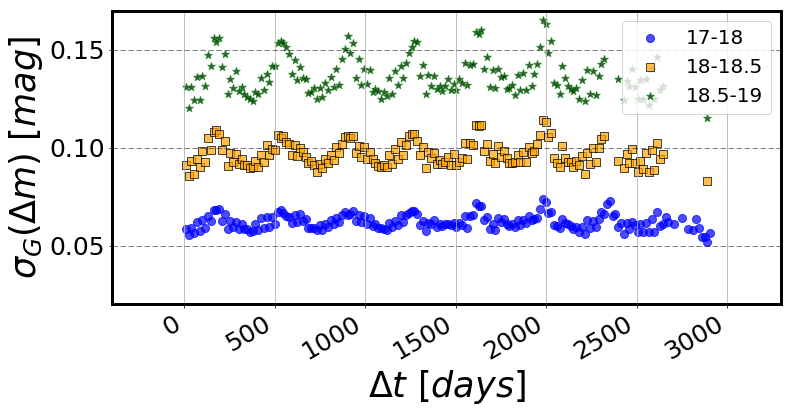}
\caption{Robust standard deviation for CRTS standard stars, showing that the oscillatory pattern persists even with fixed number of points per bin. We combine the `blue' and `red' subsamples  (-1$<g-i<$3), yielding 5788, 3487 and 3823 stars in SDSS $r$-magnitude bins {\it bright} (green stars)  {\it medium} (orange squares)  and {\it faint} (blue circles), respectively (see Table~\ref{tab:object_count} for counts in individual subsamples). For each $\Delta t$  bin, we randomly select  20$\,$000 $\Delta m$  points. If there are less than 20$\,$000 points in a bin, we do not plot anything (this affects less than 35 bins per magnitude bin, mostly towards longer time-scales). It illustrates that the wiggles are purely due to  seasonal differences, and possibly hidden zero-point errors, unaccounted for in the CSS pipeline. This pattern does not change our overall conclusions. 
\label{fig:CRTSwiggles}} 
\end{figure}

We saw  an oscillatory pattern on plots of SF and standard deviation using CRTS data on Figs.~\ref{fig:2}, \ref{fig:4} and \ref{fig:5}. We ruled out any astrophysical origin since the effect also persisted when using only standard stars. Despite an anti-correlation of the pattern with the number of points per bin, we ruled out the statistical origin by fixing the number of points per bin.  Fig.~\ref{fig:CRTSwiggles} shows that wiggles persists even if we set the number of points per $\Delta t$  bin to 20$\,$000. We see the effect when points are separated by (2k+1)/2 yr, with $k$=0,1,2 ... . We conclude that this variation is related to the airmass which fluctuates seasonally, which was  not properly accounted for in the CSS calibration process. This is because the primary aim of CSS was to detect moving objects, which  requires only intranight consistency, and not long-term accuracy \citep{drake2013}.

\bsp	
\label{lastpage}
\end{document}